\documentclass[aps,prb,english,superscriptaddress,twocolumn,floatfix]{revtex4}
\usepackage[T1]{fontenc}
\usepackage[latin9]{inputenc}
\usepackage[a4paper]{geometry}
\usepackage{bm}
\usepackage{amstext}
\usepackage{graphicx}
\usepackage{color}

\makeatletter
\@ifundefined{textcolor}{}
{%
 \definecolor{BLACK}{gray}{0}
 \definecolor{WHITE}{gray}{1}
 \definecolor{RED}{rgb}{1,0,0}
 \definecolor{GREEN}{rgb}{0,1,0}
 \definecolor{BLUE}{rgb}{0,0,1}
 \definecolor{CYAN}{cmyk}{1,0,0,0}
 \definecolor{MAGENTA}{cmyk}{0,1,0,0}
 \definecolor{YELLOW}{cmyk}{0,0,1,0}
 }

\usepackage{textcomp}
\usepackage{lmodern}
\usepackage[T1]{fontenc}

\newcommand{\murm}{\hbox{\textmu}}
\newcommand{\muSR}{\murm $^{+}$SR}

\makeatother

\usepackage{babel}

\begin{document}

\title{Low-moment magnetism in the double perovskites
  Ba$_2$\emph{M}OsO$_{\text{6}}$ (\emph{M}=Li,Na)}

\author{Andrew~J.~Steele}
\affiliation{Oxford University Department of Physics, 
Clarendon Laboratory, Parks Road, Oxford OX1 3PU, United Kingdom}
\author{Peter~J.~Baker}
\affiliation{ISIS Pulsed Neutron and Muon Source, Science and Technology
Facilities Council, Rutherford Appleton Laboratory, Didcot OX11 0QX,
United Kingdom}
\author{Tom~Lancaster}
\affiliation{Oxford University Department of Physics, 
Clarendon Laboratory, Parks Road, Oxford OX1 3PU, United Kingdom}
\author{Francis~L.~Pratt}
\affiliation{ISIS Pulsed Neutron and Muon Source, Science and Technology
Facilities Council, Rutherford Appleton Laboratory, Didcot OX11 0QX,
United Kingdom}
\author{Isabel~Franke}
\author{Saman~Ghannadzadeh}
\author{Paul~A.~Goddard}
\author{William~Hayes}
\author{D.~Prabhakaran}
\affiliation{Oxford University Department of Physics, 
Clarendon Laboratory, Parks Road, Oxford OX1 3PU, United Kingdom}
\author{Stephen~J.~Blundell}
\altaffiliation{Corresponding author: s.blundell@physics.ox.ac.uk}
\affiliation{Oxford University Department of Physics, 
Clarendon Laboratory, Parks Road, Oxford OX1 3PU, United Kingdom}

\date{\today}

\begin{abstract}
The magnetic ground states of the isostructural double perovskites
Ba$_{\text{2}}$NaOsO$_{\text{6}}$ and
Ba$_{\text{2}}$LiOsO$_{\text{6}}$ are investigated with muon-spin
relaxation.  In Ba$_{\text{2}}$NaOsO$_{\text{6}}$ long-range magnetic
order is detected via the onset of a spontaneous muon-spin precession
signal below $T_{\mathrm{c}}=7.2\pm 0.2\;\mathrm{K}$, while in
Ba$_{\text{2}}$LiOsO$_{\text{6}}$ a static but spatially-disordered
internal field is found below 8~K.  A novel probabilistic argument is
used to show from the observed precession frequencies that the
magnetic ground state in Ba$_{\text{2}}$NaOsO$_{\text{6}}$ is most
likely to be low-moment ($\approx 0.2$\,$\mu_{\rm B}$) ferromagnetism
and not canted antiferromagnetism.  
Ba$_2$LiOsO$_6$ is antiferromagnetic and we find a spin-flop
transition at 5.5\,T.
A reduced osmium moment is common
to both compounds, probably arising from a combination of spin-orbit
coupling and frustration.
\end{abstract}

\pacs{999}

\maketitle 

\section{Introduction}

Metal oxides containing 5d transition metal ions provide a wealth of
novel magnetic behavior in which orbital, charge and spin degrees of
freedom play a r\^ole. The varied properties observed are due
in part to the extended character of the 5d orbitals, reducing the
effects of electron correlation. Moreover, the larger spin-orbit
coupling in 5d oxides contributes to a balance of competing energy
scales substantially different from that in the more familiar 3d
transition metal oxides. In this context, osmium compounds provide a
number of interesting examples. Though OsO$_{\text{2}}$,
SrOsO$_{\text{3}}$ and BaOsO$_{\text{3}}$ [all containing
  Os$^{\text{4+}}$ (5d$^{\text{4}}$)] are Pauli
paramagnets~\cite{Greedan1968}, double perovskites of the form
Ba$_{\text{2}}$(Na,~Li)OsO$_{\text{6}}$ [containing
  Os$^{\text{7+}}$ (5d$^{\text{1}}$), with ions arranged on a
  face-centred cubic (fcc) lattice] exhibit Mott-insulating,
$S=\frac{1}{2}$ local-moment behaviour
\cite{Stitzer2002-Ba2NaOsO6-growth,Stitzer2003}.  However, the
insulator barium sodium osmate (Ba$_{2}$NaOsO$_{6}$) has drawn
particular attention~\cite{Erickson2007-Ba2NaOsO6-PRL} due to its
seemingly contradictory combination of negative Weiss temperature
($\approx-10\;\mathrm{K}$) and yet weak ferromagnetic moment
($\approx0.2\mu_{\mathrm{B}}\mathrm{/formula\; unit}$) below
$T_{\mathrm{c}}\approx7\;\mathrm{K}$. This contrasts with the
isostructural Ba$_{\text{2}}$LiOsO$_{\text{6}}$, which has both a
negative Weiss temperature and no ferromagnetic moment
\cite{sleight,Stitzer2002-Ba2NaOsO6-growth}.  The high-temperature
paramagnetic moment $\mu_{\rm eff}\approx 0.6\mu_{\rm B}$ in
Ba$_{2}$NaOsO$_{6}$ is also indicative of substantial spin-orbit
coupling \cite{Erickson2007-Ba2NaOsO6-PRL}.

\begin{figure}
\includegraphics[width=8cm]{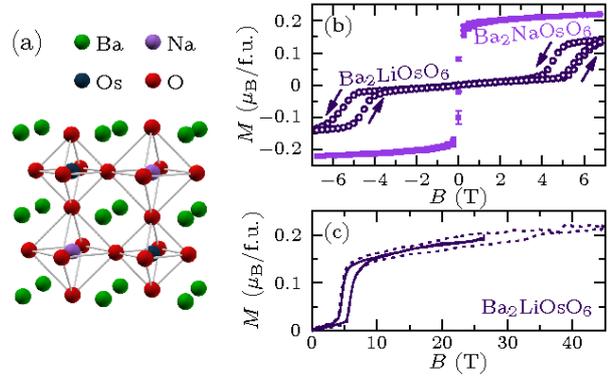}
  \caption{ (a) Double-perovskite crystal structure.  Alternation of
    OsO$_6$ octahedra (top left and bottom right) and NaO$_6$
    octahedra, with barium ions occupying the remaining space.  (b)
    Magnetization measured at 2\,K in a SQUID magnetometer for
    Ba$_2$\emph{M}OsO$_{\text{6}}$.  (c) Magnetization measured at
    4\,K in pulsed field for Ba$_2$LiOsO$_6$, calibrated
    against the SQUID magnetization data. Data from a 45\,T shot
    (dashed line) are noisier than those from a 25\,T shot (solid line)
    but show no evidence for additional magnetic transitions.
\label{fig:Ba2NaOsO6-structure}}
\end{figure}

The double-perovskite structure of Ba$_{\text{2}}$NaOsO$_{\text{6}}$
is shown in Fig.~\ref{fig:Ba2NaOsO6-structure}(a); note that
sodium and osmium ions inhabit alternate oxygen octahedra.  The
negative Weiss constant raises the possibility that the material is a
canted antiferromagnet, although first-principles density functional
theory electronic structure calculations indicate that a ferromagnetic
state has lower energy~\cite{Xiang2007-Ba2NaOsO6-PRB}.  The large
hybridization between the osmium 5d and oxygen 2p orbitals leads to a
very large crystal field splitting (several eV) between the $e_{g}$
and $t_{2g}$ bands.  It has been suggested that the low moment 
arises from a partial cancellation of orbital and spin angular momenta
in the occupied spin-up $t_{2g}$ band~\cite{Xiang2007-Ba2NaOsO6-PRB}.
A ferromagnetic phase in ordered double perovskites (with moments
along [110]) is also predicted from a  mean-field treatment of the
orbital-dependent exchange for certain values of parameters in a
model Hamiltonian \cite{chen2010}.
An alternative mechanism~\cite{Lee2007-Ba2NaOsO6-EPL} involves the
frustration of both the antiferromagnetic interactions and the orbital
ordering by the fcc lattice, resulting in a delicate balance of
interactions which favors ferromagnetism for
Ba$_{\text{2}}$NaOsO$_{\text{6}}$, but antiferromagnetism for
Ba$_{\text{2}}$LiOsO$_{\text{6}}$.
In view of these competing explanations, further experimental data
which can distinguish between different magnetic configurations are
desirable. We have used muon-spin relaxation (\muSR ) as a local
probe of magnetism in Ba$_{\text{2}}$NaOsO$_{\text{6}}$, and find that our data
are consistent with the development of long-range ferromagnetic order
with a reduced moment.  A similarly reduced moment is likely for
Ba$_2$LiOsO$_6$.

\section{Experimental}

Crystallites of Ba$_2$NaOsO$_6$ and Ba$_2$LiOsO$_6$ were grown using a
flux method \cite{Stitzer2002-Ba2NaOsO6-growth}.  Powders of Os
(99.8\%), Ba(OH)$_2\cdot$8H$_2$O (98\%) and high-purity
NaOH$\cdot$H$_2$O (99.996\%) or LiOH (99.995\%) and KOH (99.99\%) were
mixed in the ratio 1:2.1:300 or 1:2.1:140:75 respectively. These
mixtures were each placed in an alumina crucible inside a thick quartz
tube which was inserted into a 600$^\circ$C pre-heated tube furnace
where it was held for 3 days.  The furnace was then rapidly cooled to
room temperature and small single crystals were harvested from each
crucible.  Our muon experiments used a very large number of these
crystals without alignment.  Magnetization 
[Fig.~\ref{fig:Ba2NaOsO6-structure}(b)] 
and susceptibility (not shown) measurements 
using a SQUID
magnetometer 
are consistent
with earlier work
\cite{Stitzer2002-Ba2NaOsO6-growth,Erickson2007-Ba2NaOsO6-PRL} but
also reveal a spin-flop transition in Ba$_2$LiOsO$_6$ at around
5.5\,T.  Pulsed field magnetometry revealed no additional
transitions up to 45\,T
[Fig.~\ref{fig:Ba2NaOsO6-structure}(c)].

In a \muSR\ experiment \cite{Blundell1999-mureview}, spin-polarized
positive muons are implanted into a sample. The muons usually stop in
sites with high electron density.  Their spins precess around the
local magnetic field with a frequency $\nu=\gamma_{\mu}\vert
B\vert/2\pi$ where $\gamma_{\mu}=2\pi\times135.5\;\mathrm{MHz\,
  T^{-1}}$ is the muon gyromagnetic ratio.  Muons are unstable with
mean lifetime $2.2\;\mathrm{\mu s}$, and decay into a positron and two
neutrinos, the former being preferentially emitted along the direction
of muon spin. Detectors record the direction of positron emission,
whose time dependence tracks the ensemble of muon spins rotating
around their respective local $\bm{B}$-fields.  In these experiments,
the detectors are divided into a forward ($\mathrm{F}$) and backward
($\mathrm{B}$) detector bank, and the direction of preferential
positron emission is represented by the asymmetry between
$N_{\mathrm{F}}(t)$ and $N_{\mathrm{B}}(t)$, the number of positrons
detected in each detector bank as a function of time. The asymmetry
function, which is proportional to the spin polarisation of the muon
ensemble, is defined as $ A(t)=[N_{\mathrm{F}}(t)-\alpha
  N_{\mathrm{B}}(t)]/[N_{\mathrm{F}}(t)+\alpha N_{\mathrm{B}}(t)]$,
where $\alpha$ is an experimentally-determined parameter dependent on
apparatus geometry and detector efficiency.
\begin{figure}
\includegraphics[width=6.5cm]{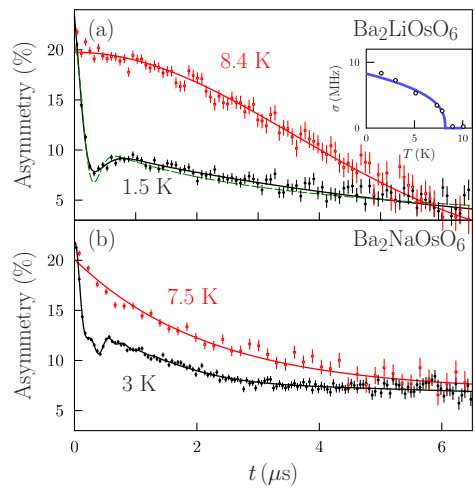}
\caption{Muon data above and below the magnetic transition for (a)
  Ba$_{\text{2}}$LiOsO$_{\text{6}}$ and (b)
  Ba$_{\text{2}}$NaOsO$_{\text{6}}$. The inset to (a) shows the
  temperature dependence of the Gaussian relaxation rate in
  Ba$_{\text{2}}$LiOsO$_{\text{6}}$ (the solid line is a best fit to a
  phenomenological $[1-(T/T_{\rm N})^\alpha)]^\beta$ form, see text).
  The 1.5\,K data in (a) can be fitted to either an
  exponentially-damped Kubo-Toyabe function (green dashed line) or an
  exponentially-damped oscillation plus a fast-relaxing component
  (black solid line).  Fits in (b) are described in the text.
  \label{fig:Ba2NaOsO6-At}}
\end{figure}

Our \muSR\ measurements were performed using the General Purpose
Surface-Muon Instrument at the Swiss Muon Source (S$\mu$S), Paul
Scherrer Institut, Switzerland.  Samples were wrapped in
$25\;\mathrm{\mu m}$ Ag foil and mounted on a `flypast' holder
comprising two silver prongs, thereby minimizing the background
signal. 

\section{Results}
\begin{figure}
\includegraphics[width=6.5cm]{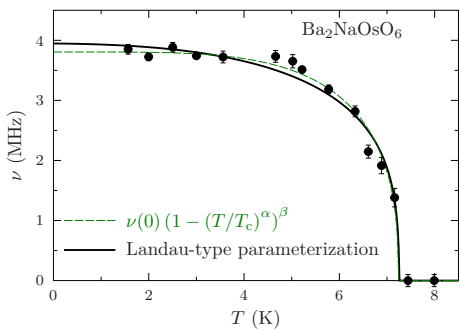}
\caption{The frequency $\nu(T)$ 
extracted from fits to $A(t)$ spectra as a function of
sample temperature $T$, along with fits to various models (see text).
Data are only
shown for the larger frequency $\nu_{1}$ because $\nu_{2}$ was held in fixed
proportion during fitting (see text). 
\label{fig:Ba2NaO6O6-nu(T)}}
\end{figure}

Typical zero-field \muSR\ spectra measured above and below the
transition temperature $T_{\mathrm{c}}$ are shown in
Fig.~\ref{fig:Ba2NaOsO6-At} for both Ba$_2$LiOsO$_6$ and
Ba$_2$NaOsO$_6$.  For Ba$_{\text{2}}$LiOsO$_{\text{6}}$, a single,
heavily-damped oscillation is present at low temperature, signifying 
a static but
spatially-disordered spontaneous field. At higher temperature the
relaxation is Gaussian, signifying the paramagnetic state.  Focussing
only on the early-time relaxation and plotting the relaxation rate
(crudely parametrizing the development of static fields) as a function
of temperature [inset to Fig.~\ref{fig:Ba2NaOsO6-At}(a)] displays the
magnetic transition at 8\,K.  For Ba$_{\text{2}}$NaOsO$_{\text{6}}$,
two damped oscillations are clearly visible at low temperature,
demonstrating the existence of a transition to a state of long-range
magnetic order. Just above the transition the relaxation is exponential [Fig.~\ref{fig:Ba2NaOsO6-At}(b)],
but becomes Gaussian at higher temperatures (not shown).
Data below the transition were fitted to
\begin{equation}
A(t)= \left[ \sum_{i=1}^2 A_i e^{-\lambda_it} \cos 2\pi\nu_it 
+ A_3 e^{-\lambda_3t} \right] + A_0 e^{-\lambda_0t},
\label{eq:Ba2NaOsO6-At}
\end{equation}
where the terms in square brackets represent muons which stop inside
the sample. These comprise two oscillatory components (reflecting
those components of the initial muon spin polarization aligned
transverse to the direction of the quasistatic local magnetic field at
the muon sites) and a non-oscillatory component. The final term in
Eq.~(\ref{eq:Ba2NaOsO6-At}) accounts for a background signal from
those muons that stop in the silver sample holder or cryostat tails.
The frequencies of the two oscillatory components were held in fixed
proportion $\eta=\nu_{2}/\nu_{1}$ during fitting (thus we will now
write $\nu_{1}\equiv \nu$ and $\nu_2 \equiv \eta\nu$).  The existence
of two oscillatory components most likely indicates two
crystallographically-similar muon sites, perhaps one nearer to the
magnetic Os ion and the other nearer the non-magnetic Na ion. The
ratio of the probabilities of stopping in the two oscillating states is
$A_{2}/A_{1}=0.19\pm0.02$.  Our fits yield $A_{3}/A_{2}=0.92\pm0.01$,
 $\nu(T \to 0)=3.9\pm0.1\;\mathrm{MHz}$ and $\eta=0.4\pm0.05$.
The relaxation rate $\lambda_{1}\approx 2\;\mathrm{MHz}$ at low
temperature
and diverges as the transition is approached from below.  The second
frequency is more strongly broadened; we find
$\lambda_{2}\approx 4\;\mathrm{MHz}$ which does not vary significantly as a
function of temperature. $\lambda_{3}$ decreases from
$1.6\pm0.1\;\mathrm{MHz}$ at $1.5\;\mathrm{K}$ towards zero at the
transition temperature $T_{\mathrm{c}}$.  
The temperature dependence
of $\nu$ was fitted to the
phenomenological form
\begin{equation}\nu(T)=\nu(0)\left[1-\left({T}/{T_{\mathrm{c}}}\right)^{\alpha}
  \right]^{\beta},
\end{equation} 
with $\beta$ fixed at 0.367, corresponding to the
3D Heisenberg model, yielding
$T_{\mathrm{c}}=7.2\pm 0.2\;\mathrm{K}$
(Fig.~\ref{fig:Ba2NaO6O6-nu(T)}).  A similarly good fit can be
obtained using a Landau-type parametrization of the equation of state
of a ferromagnet [using Eq.~(18) of Kuz'min \cite{kuzmin} with
parameters appropriate for Fe], yielding the same estimate of $T_{\rm
  c}$.  The agreement with these latter two models  
is consistent with three-dimensional ferromagnetic order.

\section{Discussion}
The data extracted from
a \muSR\ experiment can be used to infer further information about
the nature of the magnetically-ordered state by examining the local
field at the muon site. Since the positive muon seeks out areas of
negative charge  density, constraints can be placed on the likely location
of stopped muons. Calculating the magnetic field at these
plausible muon sites and comparing it to 
$|\bm{B}|=2\pi\nu_{i}(0)/\gamma_{\mu}$, allows the consistency
of a suggested magnetic structure and experiment to be checked.
The magnetic field 
$\bm{B}_{\rm local}$ at the muon site is given by
\begin{equation}
\bm{B}_{\mathrm{local}}=\bm{B}_{0}+\bm{B}_{\mathrm{dipole}}+\bm{B}_{\mathrm{L}}+\bm{B}_{\mathrm{demag}}+\bm{B}_{\mathrm{hyperfine}},\end{equation}
where $\bm{B}_{0}$ represents the applied field (zero in our
experiments), $\bm{B}_{\mathrm{dipole}}$ is the dipolar field from
magnetic ions, $\bm{B}_{\mathrm{L}}=\mu_{0}\bm{M}/3$ is the Lorentz
field,
$\bm{B}_{\mathrm{demag}}$ is the demagnetizing field from the sample
surface and $\bm{B}_{\mathrm{hyperfine}}$ is the contact hyperfine
field caused by any spin density overlapping with the muon
wavefunction.  The dipolar field $\bm{B}_{\mathrm{dipole}}$ is a
function of the muon site $\bm{r}_{\mu}$ and is composed of the vector
sum of the fields from each of the magnetic ions in the crystal, so
that
\begin{equation}
\bm{B}_{\rm dipole}(\bm{r}_{\mu})=\mu \sum_{i}\frac{\mu_{0}}{4\pi r^{3}}\left[3(\hat{\bm{\mu}}_{i}\cdot\hat{\bm{r}})\hat{\bm{r}}-\hat{\bm{\mu}}_{i}\right],\end{equation}
where $\bm{r}=\bm{r}_{i}-\bm{r}_{\mu}$
is the relative position of the muon and the $i{\text{th}}$ ion 
with magnetic moment
$\bm{\mu}_{i} = \mu \hat{\bm{\mu}}_i$, and the sum is evaluated over a
large sphere centred at $\bm{r}_\mu$.

Dipole field simulations were performed for
Ba$_{\text{2}}$NaOsO$_{\text{6}}$ using a variety of magnetic models:
ferromagnetism and antiferromagnetism (ignoring any canting) with
moments orientated along $[0\,0\,1]$, $[0\,1\,1]$ and $[1\,1\,1]$.
The muon site is not known and so we adopt a probabilistic
approach~\cite{Steele2011,Blundell2009-dipole-fields}.  The metal ions in this
system are all positively charged, meaning that the muon is unlikely
to stop near them. In other oxides, muons have been shown to stop
around $0.1\;\mathrm{nm}$ from an O$^{\text{2}-}$
ion~\cite{Brewer1991-muon-1A-from-O}.  In our calculations, positions
in the unit cell were generated at random and dipole fields calculated
at sites which were both approximately this distance from an oxygen
ion ($0.09\leq r_{\mu\text{--}\mathrm{O}}\leq0.11\;\mathrm{nm}$) and
not too close to a positive ion
($r_{\mu\text{--}+}\geq0.1\;\mathrm{nm}$).  The magnitudes of the
resulting fields were then converted into muon precession frequencies,
and the resulting histogram yields the probability density function
(pdf) $f(\nu/\mu)$, evaluated as a function of precession frequency
$\nu$ divided by osmium moment $\mu$ (since the precession frequency
scales with the osmium moment).  This analysis ignores contributions
from the Lorentz field and demagnetizing field, though these cancel
each other to some extent and we estimate them to be less than 2\,MHz.
More difficult to estimate is the importance of the contact hyperfine
field, $\bm{B}_{\mathrm{hyperfine}}$, which is neglected.  Since the
Os$^{7+}$ ion is both small and positively charged, this contribution should
not be significant and would be even further reduced if $\mu$ is low.

\begin{figure}[!t]
\includegraphics[width=7.5cm]{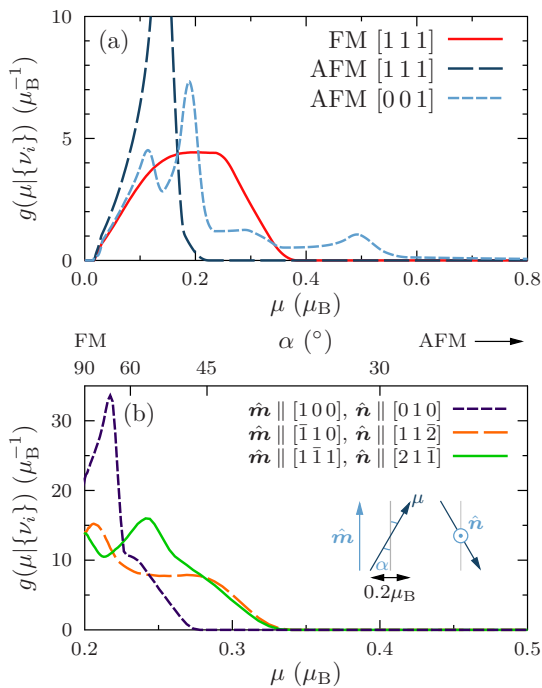}
\caption{(a) Probability density functions (pdfs) of the moment
  ($\mu$) on the Os site in Ba$_2$NaOsO$_6$ inferred by Bayes' theorem
  from dipole-field simulations using representative trial magnetic
  structures (FM=ferromagnetic, AFM=antiferromagnetic) 
  and the experimentally-observed frequencies.  (b) Pdfs 
  evaluated using the constraint of a fixed ferromagnetic component
  $\mu\sin\alpha=0.2$\,$\mu_{\rm B}$. Inset shows the assumed structure:
  antiferromagnetic moments along $\hat{\bm m}$ rotated as shown by an angle
  $\alpha$ about $\hat{\bm n}$.  }
\label{fig:Ba2NaOsO6-nu-pdfs}
\end{figure}

Since $\nu$ is obtained from experiment, what we would like to know is
$g(\mu\vert\nu)$, the pdf of $\mu$ {\it given} the
observed $\nu$.  This can be obtained from our calculated $f(\nu/\mu)$
using Bayes' theorem \cite{sivia}, which yields
\begin{equation}
g(\mu\vert\nu) = { { 1 \over \mu} f(\nu/\mu) \over \int_0^{\mu_{\rm max}} 
{1 \over \mu'} f(\nu/\mu')\,{\rm d}\mu' }, 
\end{equation}
where we have assumed a prior probability for the osmium moment that
is uniform between zero and $\mu_{\rm max}$.  We take $\mu_{\rm
  max}=1\,\mu_{\rm B}$, although our results are
insensitive to the precise value of $\mu_{\rm max}$ as long as it is
reasonably large.  When multiple frequencies $\nu_{i}$ are present in
the spectra, it is necessary to multiply their probabilities of
observation in order to obtain the chance of their simultaneous
observation, so we evaluate $g(\mu\vert\{\nu_i\}) \propto \prod_{i}
\int_{\nu_i-\Delta\nu_i}^{\nu_i+\Delta\nu_i}f(\nu_i/\mu)\,{\rm
  d}\nu_i$, where $\Delta\nu_i$ is the error on the fitted frequency.

The results of this are shown in Fig.~\ref{fig:Ba2NaOsO6-nu-pdfs}(a),
and show that for all collinear magnetic structures considered a low
$\mu$ is overwhelmingly likely. Furthermore, for the
ferromagnetic case the most probable $\mu\approx 0.2$\,$\mu_{\rm B}$,
consistent with the magnetization measurements (the AFM structures are
also consistent with a low $\mu$, but are ruled out by the
magnetization measurements).  An alternative method
of testing the hypothesis of a reduced moment is to consider possible
canted structures in which moments are counter-rotated by an angle
$\alpha$, subject to the constraint that
$\mu\sin\alpha=0.2$\,$\mu_{\rm B}$.  The results are shown in
Fig.~\ref{fig:Ba2NaOsO6-nu-pdfs}(b) and are again consistent with a
reduced moment $\mu$, certainly lower than the high-temperature
paramagnetic moment.

Thus, this probabilistic methodology suggests that the small magnetic
moment of ordered Ba$_{\text{2}}$NaOsO$_{\text{6}}$ is likely to be
true weak-moment ferromagnetism, rather than an artifact of a canted
magnetic structure with larger moments.  The moments in
Ba$_2$LiOsO$_6$ appear to be static but show a larger degree of
disorder than those of the Ba$_{\text{2}}$NaOsO$_{\text{6}}$ for
$T<T_{\rm N}$.  The local field at the muon site (estimated from
either the extracted Kubo-Toyabe field distribution width or the
damped oscillation frequency) is $\sim 0.01$\,T.  A similar Bayesian
analysis (not shown) also yields a likely low moment $\lesssim
0.2$\,$\mu_{\rm B}$, consistent with that measured in the spin-flop
state [Fig.~\ref{fig:Ba2NaOsO6-structure}(b,c)].  Thus a low moment is
common to both compounds.  In contrast to BaIrO$_3$, which achieves a
low ferromagnetic moment due to charge-density wave formation
\cite{Brooks2005}, in these compounds the origin must be rather
different.  It has been
postulated~\cite{Xiang2007-Ba2NaOsO6-PRB,Lee2007-Ba2NaOsO6-EPL} that
the d$^{\text{1}}$ spin moment of Os is compensated by the $t_{2g}$
$L=1$ orbital moment induced by the very strong spin-orbit coupling.
Partial orbital quenching by the environment is thought to destroy the
perfect compensation of the moment, leading to a small remaining
magnetic moment.  However, an important
complication~\cite{chen2010,Lee2007-Ba2NaOsO6-EPL} is that the
antiferromagnetic orbital-dependent coupling between OsO$_6$ clusters
takes place on an fcc lattice. It is therefore strongly frustrated and
may hence be responsible for the sensitivity of the ground state to
minor chemical changes, explaining the difference between the magnetic
properties of isostructural Ba$_2$LiOsO$_6$ and Ba$_2$NaOsO$_6$.

\acknowledgments
Part of this work was carried out at S\murm S, 
Paul Scherrer Institut, Villigen, Switzerland and we are grateful to
A. Amato for technical support. This work is supported by the EPSRC,
UK.

\bibliographystyle{apsrev}

\begin{thebibliography}{14}
\expandafter\ifx\csname natexlab\endcsname\relax\def\natexlab#1{#1}\fi
\expandafter\ifx\csname bibnamefont\endcsname\relax
  \def\bibnamefont#1{#1}\fi
\expandafter\ifx\csname bibfnamefont\endcsname\relax
  \def\bibfnamefont#1{#1}\fi
\expandafter\ifx\csname citenamefont\endcsname\relax
  \def\citenamefont#1{#1}\fi
\expandafter\ifx\csname url\endcsname\relax
  \def\url#1{\texttt{#1}}\fi
\expandafter\ifx\csname urlprefix\endcsname\relax\def\urlprefix{URL }\fi
\providecommand{\bibinfo}[2]{#2}
\providecommand{\eprint}[2][]{\url{#2}}

\bibitem[{\citenamefont{Greedan et~al.}(1968)\citenamefont{Greedan, Willson,
  and Haas}}]{Greedan1968}
\bibinfo{author}{\bibfnamefont{J.~E.} \bibnamefont{Greedan}},
  \bibinfo{author}{\bibfnamefont{D.~B.} \bibnamefont{Willson}},
  \bibnamefont{and} \bibinfo{author}{\bibfnamefont{T.~E.} \bibnamefont{Haas}},
  \bibinfo{journal}{Inorg. Chem.} \textbf{\bibinfo{volume}{11}},
  \bibinfo{pages}{2461} (\bibinfo{year}{1968}).

\bibitem[{\citenamefont{Stitzer et~al.}(2002)\citenamefont{Stitzer, Smith, and
  zur Loye}}]{Stitzer2002-Ba2NaOsO6-growth}
\bibinfo{author}{\bibfnamefont{K.~E.} \bibnamefont{Stitzer}},
  \bibinfo{author}{\bibfnamefont{M.~D.} \bibnamefont{Smith}}, \bibnamefont{and}
  \bibinfo{author}{\bibfnamefont{H.-C.} \bibnamefont{zur Loye}},
  \bibinfo{journal}{Solid State Sciences} \textbf{\bibinfo{volume}{4}},
  \bibinfo{pages}{311} (\bibinfo{year}{2002}).

\bibitem[{\citenamefont{Stitzer et~al.}(2003)\citenamefont{Stitzer, Abed,
  Smith, Davis, Kim, Darriet, and zur Loye}}]{Stitzer2003}
\bibinfo{author}{\bibfnamefont{K.~E.} \bibnamefont{Stitzer}},
  \bibinfo{author}{\bibfnamefont{A.~E.} \bibnamefont{Abed}},
  \bibinfo{author}{\bibfnamefont{M.~D.} \bibnamefont{Smith}},
  \bibinfo{author}{\bibfnamefont{M.~J.} \bibnamefont{Davis}},
  \bibinfo{author}{\bibfnamefont{S.-J.} \bibnamefont{Kim}},
  \bibinfo{author}{\bibfnamefont{J.}~\bibnamefont{Darriet}}, \bibnamefont{and}
  \bibinfo{author}{\bibfnamefont{H.-C.} \bibnamefont{zur Loye}},
  \bibinfo{journal}{Inorg. Chem.} \textbf{\bibinfo{volume}{42}},
  \bibinfo{pages}{947} (\bibinfo{year}{2003}).

\bibitem[{\citenamefont{Erickson et~al.}(2007)\citenamefont{Erickson, Misra,
  Miller, Gupta, Schlesinger, Harrison, Kim, and
  Fisher}}]{Erickson2007-Ba2NaOsO6-PRL}
\bibinfo{author}{\bibfnamefont{A.~S.} \bibnamefont{Erickson}},
  \bibinfo{author}{\bibfnamefont{S.}~\bibnamefont{Misra}},
  \bibinfo{author}{\bibfnamefont{G.~J.} \bibnamefont{Miller}},
  \bibinfo{author}{\bibfnamefont{R.~R.} \bibnamefont{Gupta}},
  \bibinfo{author}{\bibfnamefont{Z.}~\bibnamefont{Schlesinger}},
  \bibinfo{author}{\bibfnamefont{W.~A.} \bibnamefont{Harrison}},
  \bibinfo{author}{\bibfnamefont{J.~M.} \bibnamefont{Kim}}, \bibnamefont{and}
  \bibinfo{author}{\bibfnamefont{I.~R.} \bibnamefont{Fisher}},
  \bibinfo{journal}{Phys. Rev. Lett.} \textbf{\bibinfo{volume}{99}},
  \bibinfo{pages}{016404} (\bibinfo{year}{2007}).

\bibitem[{\citenamefont{Sleight et~al.}(1962)\citenamefont{Sleight, Longo, and
  Ward}}]{sleight}
\bibinfo{author}{\bibfnamefont{A.~W.} \bibnamefont{Sleight}},
  \bibinfo{author}{\bibfnamefont{J.}~\bibnamefont{Longo}}, \bibnamefont{and}
  \bibinfo{author}{\bibfnamefont{R.}~\bibnamefont{Ward}},
  \bibinfo{journal}{Inorg. Chem.} \textbf{\bibinfo{volume}{1}},
  \bibinfo{pages}{245} (\bibinfo{year}{1962}).

\bibitem[{\citenamefont{Xiang and Whangbo}(2007)}]{Xiang2007-Ba2NaOsO6-PRB}
\bibinfo{author}{\bibfnamefont{H.~J.} \bibnamefont{Xiang}} \bibnamefont{and}
  \bibinfo{author}{\bibfnamefont{M.-H.} \bibnamefont{Whangbo}},
  \bibinfo{journal}{Phys. Rev. B} \textbf{\bibinfo{volume}{75}},
  \bibinfo{pages}{052407} (\bibinfo{year}{2007}).

\bibitem{chen2010}
G. Chen, R. Pereira and L. Balents,
Phys. Rev. {\bf 82}, 174440 (2010).

\bibitem[{\citenamefont{Lee and Pickett}(2007)}]{Lee2007-Ba2NaOsO6-EPL}
\bibinfo{author}{\bibfnamefont{K.-W.} \bibnamefont{Lee}} \bibnamefont{and}
  \bibinfo{author}{\bibfnamefont{W.~E.} \bibnamefont{Pickett}},
  \bibinfo{journal}{Europhys. Lett.} \textbf{\bibinfo{volume}{80}},
  \bibinfo{pages}{37008} (\bibinfo{year}{2007}).

\bibitem[{\citenamefont{Blundell}(1999)}]{Blundell1999-mureview}
\bibinfo{author}{\bibfnamefont{S.~J.} \bibnamefont{Blundell}},
  \bibinfo{journal}{Contemp. Phys.} \textbf{\bibinfo{volume}{40}},
  \bibinfo{pages}{175} (\bibinfo{year}{1999}).

\bibitem[{\citenamefont{Kuz'min}(2008)}]{kuzmin}
\bibinfo{author}{\bibfnamefont{M.~D.} \bibnamefont{Kuz'min}},
  \bibinfo{journal}{Phys. Rev. B} \textbf{\bibinfo{volume}{77}},
  \bibinfo{pages}{184431} (\bibinfo{year}{2008}).
  
\bibitem[{\citenamefont{Steele et~al.}(2011)\citenamefont{Steele, Lancaster,
  Blundell, Baker, Pratt, Baines et~al.}}]{Steele2011}
\bibinfo{author}{\bibfnamefont{A.~J.} \bibnamefont{Steele}},
\bibinfo{author}{\bibfnamefont{T.}~\bibnamefont{Lancaster}},
  \bibinfo{author}{\bibfnamefont{S.~J.} \bibnamefont{Blundell}},
  \bibinfo{author}{\bibfnamefont{P.~J.}~\bibnamefont{Baker}},
  \bibinfo{author}{\bibfnamefont{F.~L.} \bibnamefont{Pratt}},
  \bibinfo{author}{\bibfnamefont{C.} \bibnamefont{Baines}},
  \bibinfo{author}{\bibfnamefont{M.~M.} \bibnamefont{Connor}},
  \bibinfo{author}{\bibfnamefont{H.~I.} \bibnamefont{Southerland}},
  \bibinfo{author}{\bibfnamefont{J.~L.} \bibnamefont{Manson}},  
  \bibnamefont{and} \bibinfo{author}{\bibfnamefont{J.~A.}
  \bibnamefont{Schlueter}}, \bibinfo{journal}{Phys. Rev. B}
  \textbf{\bibinfo{volume}{84}}, \bibinfo{pages}{064412}
  (\bibinfo{year}{2011}).

\bibitem[{\citenamefont{Blundell}(2009)}]{Blundell2009-dipole-fields}
\bibinfo{author}{\bibfnamefont{S.~J.} \bibnamefont{Blundell}},
  \bibinfo{journal}{Physica B} \textbf{\bibinfo{volume}{404}},
  \bibinfo{pages}{581} (\bibinfo{year}{2009}).

\bibitem[{\citenamefont{Brewer et~al.}(1991)\citenamefont{Brewer, Kiefl,
  Carolan, Dosanjh, Hardy, Kreitzman, Li, Riseman, Schleger, Zhou
  et~al.}}]{Brewer1991-muon-1A-from-O}
\bibinfo{author}{\bibfnamefont{J.}~\bibnamefont{Brewer}},
  \bibinfo{author}{\bibfnamefont{R.}~\bibnamefont{Kiefl}},
  \bibinfo{author}{\bibfnamefont{J.}~\bibnamefont{Carolan}},
  \bibinfo{author}{\bibfnamefont{P.}~\bibnamefont{Dosanjh}},
  \bibinfo{author}{\bibfnamefont{W.}~\bibnamefont{Hardy}},
  \bibinfo{author}{\bibfnamefont{S.}~\bibnamefont{Kreitzman}},
  \bibinfo{author}{\bibfnamefont{Q.}~\bibnamefont{Li}},
  \bibinfo{author}{\bibfnamefont{T.}~\bibnamefont{Riseman}},
  \bibinfo{author}{\bibfnamefont{P.}~\bibnamefont{Schleger}},
  \bibinfo{author}{\bibfnamefont{H.}~\bibnamefont{Zhou}}, \bibnamefont{et~al.},
  \bibinfo{journal}{Hyp. Int.} \textbf{\bibinfo{volume}{63}},
  \bibinfo{pages}{177} (\bibinfo{year}{1991}).

\bibitem[{\citenamefont{Sivia and Skilling}(2006)}]{sivia}
\bibinfo{author}{\bibfnamefont{D.~S.} \bibnamefont{Sivia}} \bibnamefont{and}
  \bibinfo{author}{\bibfnamefont{J.}~\bibnamefont{Skilling}},
  \emph{\bibinfo{title}{Data Analysis: A Bayesian Tutorial}}
  (\bibinfo{publisher}{OUP}, \bibinfo{address}{Oxford}, \bibinfo{year}{2006}),
  \bibinfo{edition}{2nd} ed.

\bibitem[{\citenamefont{Brooks et~al.}(2005)\citenamefont{Brooks, Blundell,
  Lancaster, Hayes, Pratt, Frampton, and Battle}}]{Brooks2005}
\bibinfo{author}{\bibfnamefont{M.~L.} \bibnamefont{Brooks}},
  \bibinfo{author}{\bibfnamefont{S.~J.} \bibnamefont{Blundell}},
  \bibinfo{author}{\bibfnamefont{T.}~\bibnamefont{Lancaster}},
  \bibinfo{author}{\bibfnamefont{W.}~\bibnamefont{Hayes}},
  \bibinfo{author}{\bibfnamefont{F.~L.} \bibnamefont{Pratt}},
  \bibinfo{author}{\bibfnamefont{P.~P.~C.} \bibnamefont{Frampton}},
  \bibnamefont{and} \bibinfo{author}{\bibfnamefont{P.~D.}
  \bibnamefont{Battle}}, \bibinfo{journal}{Phys. Rev. B}
  \textbf{\bibinfo{volume}{71}}, \bibinfo{pages}{R220411}
  (\bibinfo{year}{2005}).

\end{thebibliography}

\end{document}